\documentclass[aps,prl,twocolumn]{revtex4-1}

\usepackage{epsfig}
\usepackage{graphicx}
\usepackage{amsmath}
\usepackage{mathtools}
\usepackage{amscd}
\usepackage{amssymb}
\usepackage{amsfonts}
\usepackage{amsthm}
\usepackage{enumitem}
\usepackage{stmaryrd}
\usepackage{tikz-cd}

\usepackage[T1]{fontenc}

\usepackage{xcolor}

\def\fig#1{Fig.\ \ref{#1}}

\def\be{\begin{equation}}
\def\ee{\end{equation}}
\def\bea{\begin{eqnarray}}
\def\eea{\end{eqnarray}}

\newcommand{\sket}[1]{|#1\rangle}

\begin{document}

\title{Hunting for Majoranas}  

\author {Ali Yazdani$^1$}
\author{Felix von Oppen$^2$}
 \author{Bertrand I. Halperin$^3$}
\author{Amir Yacoby$^3$}
\affiliation{$^1$ Joseph Henry Laboratories and Department of Physics, Princeton University, Princeton, New Jersey 08540, USA}
\affiliation{$^2$ Dahlem Center for Complex Quantum Systems and Fachbereich Physik, Freie Universität Berlin, 14195 Berlin, Germany}  
\affiliation{$^3$ Department of Physics, Harvard University, Cambridge, Massachusetts 02138, USA} 

\date{\today}


\begin{abstract}

Over the last decade, there have been considerable efforts to observe non-abelian quasi-particles in novel quantum materials and devices. These efforts are motivated by the goals of demonstrating quantum statistics of quasi-particles beyond those of fermions and bosons and of establishing the underlying science for the creation of topologically protected quantum bits. In this review, we  focus on efforts to create topological superconducting phases hosting Majorana zero modes. 
We consider the lessons learned from existing experimental efforts, which are motivating both improvemensts to current platforms and exploration of new approaches. Although the experimental detection of non-abelian quasi-particles remains challenging, the knowledge gained thus far and the opportunities ahead offer high potential for discovery and advances in this exciting area of quantum physics.
\end{abstract} 

\maketitle
\narrowtext

\subsection{Introduction} 

The last decade has witnessed considerable progress towards the creation of new quantum technologies.  The recent efforts \cite{Arute2019} to demonstrate computational quantum advantage \cite{Preskill2012} using processors based on superconducting quantum bits (qubits) and efforts to simulate complex quantum phases of matter with such technologies exemplify the remarkable advances of a rapidly evolving field. The current qubit technologies, based on conventional superconductors \cite{Kjaergaard2020}, semiconductors \cite{Vandersypen2019}, trapped ions \cite{Monroe2021}, or Rydberg atoms \cite{Browaeys2020}, will undoubtedly make significant progress in the years ahead. 

Beyond these current technologies, there exist blueprints for topological qubits that leverage conceptually different physics for improved qubit performance \cite{Kitaev2003,NayakSSFD08,BeenakkerReview,AliceaReview,OregOppenAnnualReview,RamonLeoPhysicsToday}. These qubits exploit the fact that quasi-particles of topological quantum states allow quantum information to be encoded and processed in a non-local manner, providing inherent protection against decoherence \cite {Kitaev2003, Freedman1998}. Although still far from being experimentally realized, the potential benefits of this approach are evident. The inherent protection against decoherence implies better scalability, while also promising a significant reduction in the number of qubits needed for error correction.

Apart from possible applications, fundamental physics furnishes ample motivation for pursuing this line of research. Topological qubits rely on topological quantum states that are predicted to host new class of quasi-particle, termed non-abelian anyons, which their quantum properties offers fundamental new way for encoding and processing quantum information. Despite numerous attempts, experimental evidence for non-abelian anyons is still limited and their non-abelian statistics remain largely a theoretical prediction. Additionally, material synthesis, experimental advances, and theoretical modeling required to realize, explore, and understand the underlying topological quantum states provide an extraordinarily rich setting for the discovery of new quantum phenomena
 
Quasi-particles with properties distinct from those of free electrons are a hallmark of condensed matter systems. A dramatic manifestation is the fractional quantum Hall (FQH) effect. Strong interactions among electrons moving in two dimensions (2D) subject to a magnetic field create topological quantum states that host anyonic quasi-particles \cite{ArovasSW84,Halperin84}. For many fractional quantum Hall states, the anyons are abelian, but some states are expected to support non-abelian anyons \cite{MooreR91}. Read and Green \cite{Read2000} proposed that the most prominent of these states, corresponding to the Landau level filling factor $\nu=5/2$ \cite{Willett87,MooreR91}, can be understood as a superconductor of composite fermions with chiral $p$-wave symmetry. Quasi-particles of this $\nu=5/2$ state realize the simplest type of non-abelian anyons known as Ising anyons. The presence of Ising anyons implies a topological degeneracy of the quantum state. The degeneracy can be associated with Majorana zero modes (MZMs) in the corresponding $p$-wave superconductor and is the foundation for  building a topological qubit.

A deeper understanding of topological quantum states has led to the realization that a far wider range of materials can host non-abelian anyons. Read and Green \cite{Read2000} already understood that MZMs exist in vortex cores of two-dimensional $p$-wave superconductors. Kitaev \cite{Kitaev2001} recognized that MZMs occur in one-dimensional $p$-wave superconductors of spinless fermions. Whereas intrinsic $p$-wave superconductors are at best rare in nature, Fu and Kane \cite{Fu2008a,Fu2008b} showed that they can be effectively realized in hybrid structures involving conventional superconductors, an idea that has stimulated a plethora of further proposals.

As we outline here, several experimental efforts observed signatures that at first appeared consistent with theoretical predictions for MZMs. Unfortunately, many of the experiments revealed a more complex reality than anticipated, allowing for multiple interpretations of the data. While this can be seen as a severe drawback or limitation of the experimental status, we believe that this is actually a desired and welcome outcome. Advances in our understanding of a physical system largely rely on discrepancies between experiment and theory. Such discrepancies suggest that the theoretical description is too simplistic, calling for modifications or extensions. Moreover, multiple possible interpretations stimulate new and sharpened experiments, which distinguish between competing theoretical interpretations. Overall, although our ability to detect and manipulate non-abelian anyons such as MZMs remains in its infancy, the landscape of materials, tools, and ideas in this area is remarkably rich. Here we provide a perspective on the progress made over the last decade, briefly reviewing the key concepts, the lessons learned from the experimental efforts thus far, and looking ahead to potential for future progress and discovery in this field.

\begin{figure}

\includegraphics[width=\linewidth]{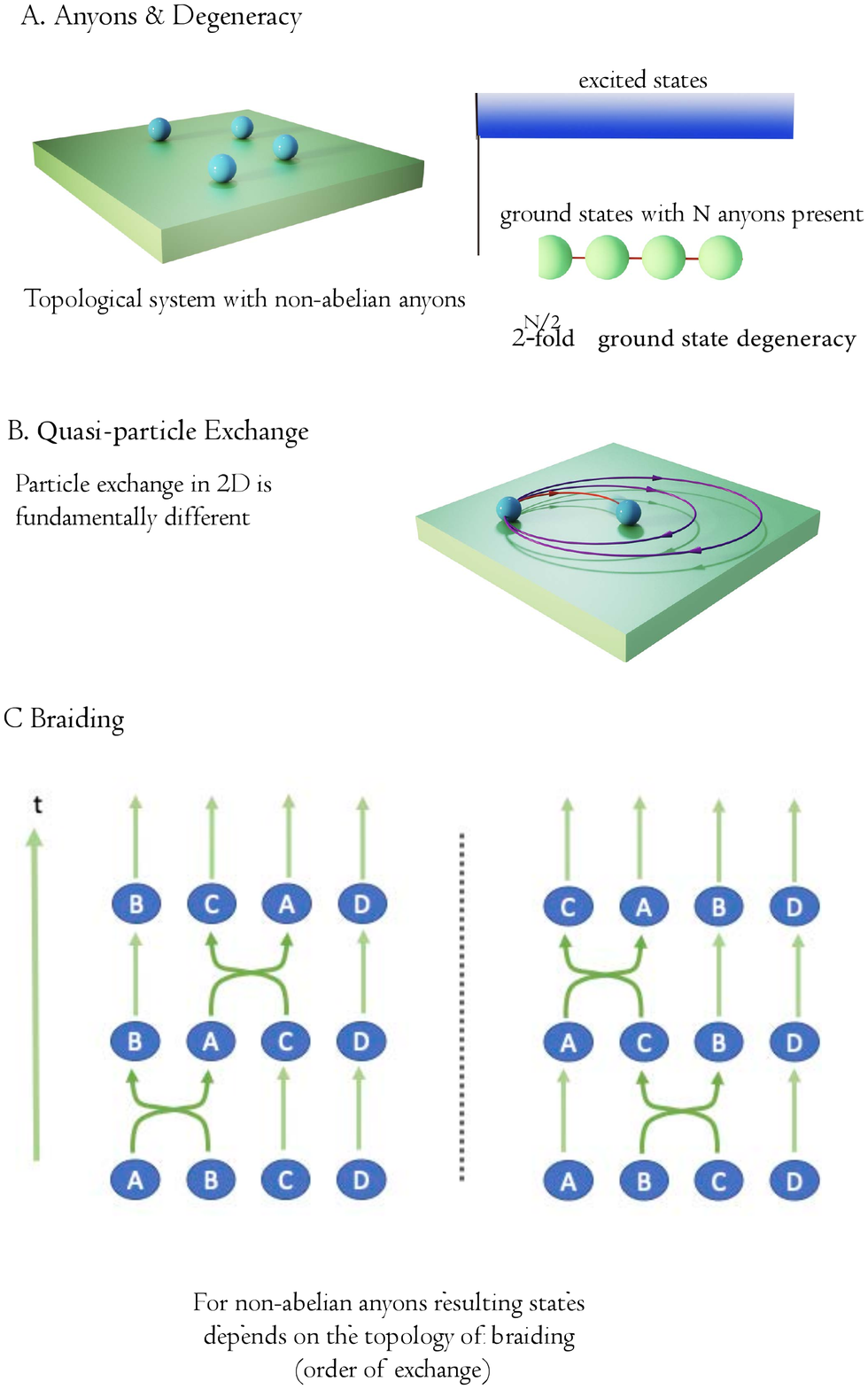}
\caption {\label{Fig1} Fundamentals. A. The ground state of a topological phase with localized anyons is highly degenerate, providing a protected subspace for storing  quantum information. The ground states are separated from excited states by an energy gap. B. The quantum exchange of quasi-particles in 2D is fundamentally different from 3D. As the figure shows the path of one particle looping, the other can keep shrinking, while in 2D the loop runs into the other particle as it shrinks. C. This difference manifests itself in topological quantum phases, which host a class of quasi-particles, termed non-abelian anyons. The low-energy quantum states of non-abelian anyons depend on the order in which exchange (or braiding) processes are performed. The state of the system after braiding is related to the initial state by a unitary transformation, which depends only on the topology of the braiding operation and can be exploited to process quantum information. The illustration shows two braiding sequences of anyons A,B,C,D (such as Ising anyons or MZMs), which implement different unitary transformations and generally result in different quantum states. 
}
\end{figure}

\subsection{Fundamentals}

Non-abelian statistics are a property of quasi-particles or defects in special two-dimensional (2D) electron systems such as FQH systems \cite{MooreR91,Wen91a,RR,Nayak1996} or topological superconductors \cite{Read2000,Ivanov2001} (\fig{Fig1}). In the absence of defects, these systems have an energy gap for excitations far from any boundaries. But when there are defects or quasi-particles at fixed locations, the ground state becomes a manifold of nearly-degenerate low-energy eigenstates, whose dimension grows exponentially with the number of defects. The energy differences between these states are predicted to fall off exponentially with the separation between the quasi-particles, so, in principle, they can be exceedingly small. In the case of the fractional quantized Hall state at filling factor $\nu=5/2$, the proposed non-abelian quasi-particles are objects with electric charge $\pm e/4$ \cite{MooreR91}, whereas in the model of a  $p_x + i p_y$ superconductor, they are associated with vortex cores (in two dimensions) \cite{Read2000} or domain walls (in one dimension) \cite{Kitaev2001}. (Here, $e$ denotes the elementary charge.)

If well-separated quasi-particles are slowly moved around each other or interchanged, in such a way that the sets of initial and final positions for each quasi-particle type are identical, the final state of the system will be related to the initial state by a unitary transformation in the Hilbert space of low-energy eigenstates. Furthermore, if this process is fast compared to the exponentially small energy splittings of the Hilbert space, but slow compared to the energy gap for other excitations, the unitary transformation will depend only on  the topology of the braiding of the quasi-particle world lines. It will be independent of other details and will be unaffected by any local perturbations. Because the resulting unitary transformation will generally depend on the order in which the interchanges have been performed, the system is  said to obey non-abelian statistics. If braiding can be performed on the appropriate time scale, the unitary transformations could be used to perform topologically protected quantum computations \cite{Kitaev2003}.  

Ising anyons, which are the simplest non-abelian quasi-particles, have ground-state degeneracy $2^{N/2}$, where $N$ is the number of quasi-particles. This can be described by associating with each localized quasi-particle a MZM, characterized by a Majorana operator $\gamma_j$ \cite{Read2000}.  Majorana operators for different quasi-particles anticommute, like ordinary fermion operators, but a single Majorana  operator obeys the relations $\gamma_j = \gamma_j^\dagger$ and $\gamma_j^2 = 1$ ({\it{i.e.}}, it is its own anti-particle).  
If $\sket{\Psi}$ belongs to the manifold of degenerate ground states, then  $\gamma_j \sket{\Psi}$ will also belong to this manifold, but have opposite fermion number parity. If quasi-particles can be moved around, their MZMs will move with them, leading to the unitary transformations within the degenerate ground-state manifold underlying non-abelian statistics. Equivalent unitary transformations can be produced without physical motion, if it is possible to effectively turn on and off the interactions between nearby MZMs \cite{Sau2011,van_Heck2012,Hyart2013,Aasen2016}. 
Alternative schemes induce or replace braiding operations by measurement protocols \cite{Bonderson2008,Karzig2017,Litinski2018,Knapp2020}.

\begin{figure}

\includegraphics[width=\linewidth]{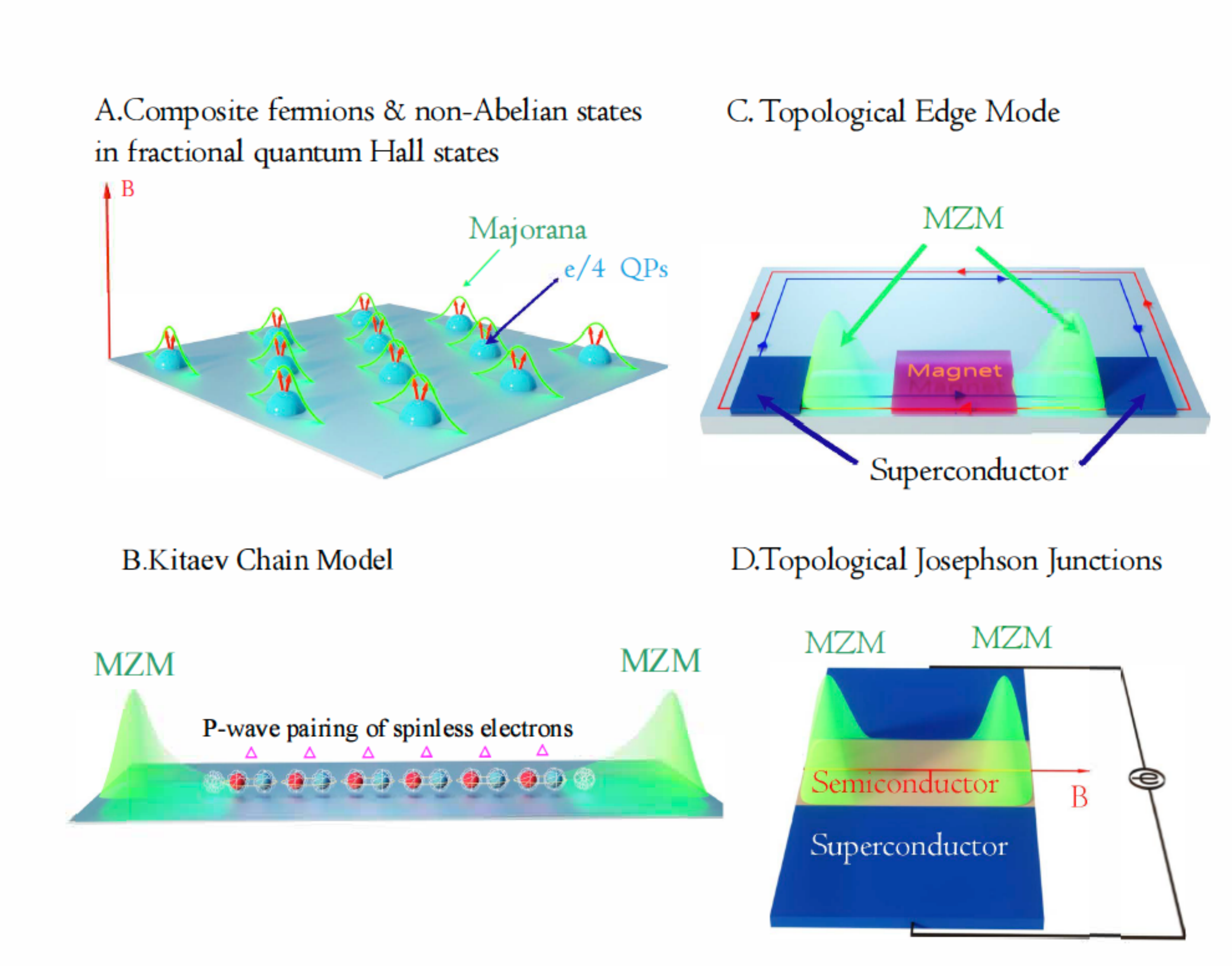}
\caption {\label{Fig2} Model Systems. There is a diverse set of possible platforms for realizing non-abelian anyons such as Majorana zero modes (MZMs). A. In some fractional quantum Hall states, such as $\nu = 5/2$, strong electronic correlations create topological phases that are predicted to support fractionally charged quasi-particles (QPs) with associated MZMs. B. Several platforms such as semiconductor quantum wires and chains of magnetic adatoms on superconducting substrates are closely related to the much-studied Kitaev chain (see also Box 1). Hopping and p-wave pairing of spinless electrons in one dimension can create the conditions for localizing isolated MZMs at the ends of the chain. The blue and red balls represent p-wave pairing amplitude ($\Delta$) that is on nearest neighbor site on a chain. C. Effective $p$-wave superconductivity can also be realized by proximity coupling  the helical edge mode (red and blue arrows at the edges) of a 2D topological insulator to a magnetic insulator and a superconductor, both of which will gap the edge mode. A MZM is localized at the boundary between these gapped regions. D. Combining a strong spin-orbit material and superconducting electrodes into a Josephson junction can also emulate the Kitaev chain. The junction hosts MZMs at the two ends, when tuned to the topological phase by adjusting the phase difference across the junction and applying a magnetic field.}
\end{figure}

\medskip
\noindent\fbox{%
    \parbox{0.47\textwidth}{%
\textbf{BOX1: Kitaev chain.} -- Kitaev's model of a spinless $p$-wave superconductor exemplifies the emergence of MZMs at the boundaries of a topological superconductor \cite{Kitaev2001}. The spinless fermions (creation operator $c_j^\dagger$ at site $j$) hop along a chain with amplitude $t$ subject to nearest-neighbor $p$-wave pairing $\Delta$,
\begin{equation}
  H= \sum_j \left\{ -t c^\dagger_{j+1} c^{\phantom{\dagger}}_j +\Delta c^\dagger_{j+1} 
   c^\dagger_j + \text{h.c.} \right\} -\mu \sum_j c^\dagger_{j} c^{\phantom{\dagger}}_j.
\end{equation}
The Kitaev chain enters a topological superconducting phase, when the chemical potential $\mu$ falls within the normal-state band $\epsilon_k=-2t\cos k$ (with $k$ denoting the fermion wave vector). Reflecting the bulk-boundary correspondence, this phase supports gapless boundary excitations at the ends of the chain, which are isolated MZMs. Apart from exponentially small corrections, a finite chain with a pair of MZMs $\gamma_1$ and $\gamma_2$ has two degenerate ground states, which differ in fermion parity and can be characterized by the occupation $d^\dagger d$ of the conventional fermion mode $d = \frac{1}{2}(\gamma_i - i \gamma_2)$.
}}
\medskip

\subsection{Model Systems}

\textbf{Fractional quantum Hall systems} The first systems that were predicted \cite{MooreR91} to host non-abelian anyons are the even-denominator FQH states observed in GaAs heterostructures at Landau-level filling fractions $\nu = 5/2$ and 7/2 \cite{Willett87}. There are, currently, three competing theoretical models for these states (see below). In all three cases, it is predicted that the elementary quasi-particles should be Ising anyons with charge $\pm e/4$. Each of these models can be mapped mathematically onto a topological superconductor interacting with an emergent gauge field, with the $e/4$ quasi-particles appearing as vortices (Fig.\ 2A).

\textbf{Kitaev's 1D model} Whereas in 2D, MZMs are associated with vortex cores, \cite{Kitaev2001} showed that MZMs also emerge at the ends of one-dimensional topological superconductors. He discusses a simple and elegant model of spinless fermions hopping on a linear chain in the presence of nearest-neighbor $p$-wave superconducting pairing $\Delta$ \cite{Kitaev2001} (see BOX 1 and Fig.\ 2B).  Although braiding of quasi-particles is impossible in a strictly 1D setting, the chains can be connected into a 2D network, which allows for braiding and non-abelian statistics \cite{Alicea2011}.

At first sight, there are severe obstacles to a physical realization of the Kitaev chain. Electrons have spin, conventional superconductors form $s$-wave rather than $p$-wave Cooper pairs, and thermal fluctuations suppress superconducting order in a one-dimensional setting. Several approaches have been proposed to overcome these challenges.

\textbf{Topological boundary modes and vortex cores} 
With the advent of topological band theory, it was predicted that boundary modes of 2D or 3D topological insulators (TI) can be used for creating topological superconductors \cite{Fu2008a,Fu2008b}. The edge mode of a 2D TI or the 2D surface state of a 3D TI exhibit spin-momentum locking, a property that can be viewed as creating effectively spinless quasi-particles, as in the Kitaev model. The edge mode of a 2D TI can be gapped out by proximity coupling to a superconductor with $s$-wave pairing or to a ferromagnetic insulator. Isolated MZMs appear at domain walls between these two types of gapped regions (Fig.\ 2C). A 2D analog can be realized by inducing superconductivity in the surface state of a 3D TI by proximity; MZM are then expected to be  bound to individual vortices induced by a perpendicular magnetic field. Individual vortices also host trivial subgap Caroli-de Gennes-Matricon states, the spectrum of which would be shifted in energy when a MZM appears at the core of the vortices. However, the MZM may be separated from finite-energy Caroli-de Gennes-Matricon states by possibly very small energies,making their detection challenging.

\textbf{Quantum wires and atomic chains} 1D topological superconductivity and MZMs can also be engineered in more conventional materials, using a by now familiar recipe \cite{Sau2010,Alicea2010,Lutchyn2010,Oreg2010}. One induces superconductivity into a (quasi) 1D system by proximity from a bulk superconductor with conventional $s$-wave pairing and suppresses the doubling of the Fermi surface by applying a magnetic field or magnetism without quenching the bulk superconductor. In the presence of spin-orbit coupling, either in the proximity-providing superconductor or in the 1D system, the spin-singlet Cooper pairs of the bulk superconductor can still proximitize the 1D system, where they effectively induce $p$-wave rather than $s$-wave pairing. As pairing in the 1D system is inherited from a bulk superconductor, the superconducting correlations are not destroyed by thermal fluctuations. Following this hybrid approach,  experimental searches for MZM have thus far focused on two classes of systems, semiconductor quantum wires and chains of magnetic adatoms on superconducting substrates.

Semiconductor quantum wires made of InAs or InSb are known to have strong spin-orbit coupling and are readily proximitized by a superconductor (Fig.\ 3B). Material science advances motivated by the search for MZMs have led to epitaxial growth of superconductors (Al) on semiconductor quantum wires with exquisite interface quality and excellent proximity coupling \cite{Krogstrup2015}. A variant to create 1D channels is to use gating of a 2D electron gas in InAs proximitized by an epitaxially grown superconductor \cite{Shabani2016}. A useful feature of expitaxial semiconductor-superconductor materials is that one can realize structures with substantial charging energies. The associated Coulomb blockade physics admits additional experimental tests of MZMs \cite{Albrecht2016,Heck2016} and is predicted to lead to Majorana teleportation \cite{Fu2010} as well as a topological Kondo effect \cite{Beri2012}. Coulomb blockade also suppresses quasi-particle poisoning by stabilizing specific charge states \cite{Albrecht2017}. This is an important ingredient in some of the most promising designs of a topological qubit (see Box 2) \cite{Karzig2017,Plugge2017,OregOppenAnnualReview}. 

Chains of magnetic adatoms on conventional superconducting substrates can also effectively emulate the Kitaev model \cite{NadjPerge2013}. This platform lends itself to in-situ characterization using high resolution scanning tunneling microscopy (STM) and spectroscopy (STS). Densely packed chains of adatoms form spin-polarized 1D bands by hybridizing their $d$-orbitals, for instance in ferromagnetically ordered chains (Fig.\ 3B). When an odd number of these bands crosses the Fermi energy of the substrate superconductor and spin-orbit coupling enables proximitizing the spin-polarized bands, a $p$-wave gap forms, effectively resulting in a 1D topological superconductor with MZMs localized at the ends of the chain \cite{NadjPerge2014, Li2014}. Alternatively, the chain can order into a spin spiral, which might self-tune the system into the topological phase \cite{Braunecker2013,Klinovaja2013,Vazifeh2013}. A characteristic feature of this platform is the strong hybridization of the adatoms with the superconducting substrate, which has been shown to result in strongly localized Majorana zero modes \cite{Peng2015a}.

Topological superconductivity has also been predicted in dilute chains, in which coupling between adatoms is entirely mediated via the superconducting substrate \cite{NadjPerge2013,Pientka2013,Steiner2022}. In this limit, each adatom induces one (or more) pairs of spin-polarized Yu-Shiba-Rusinov (YSR) states symmetric in energy about the center of the superconducting gap of the substrate. If the adatoms order magnetically, the subgap states hybridize along the chain and form YSR bands. Topological superconductivity ensues when the bands emerging from the positive- and negative-energy YSR states overlap and develop a gap due to induced $p$-wave superconductivity. For ferromagnetic order, such a gap requires the presence of spin-orbit coupling. 

\textbf{Josephson junctions} Josephson junctions coupling two conventional superconductors via a semiconductor with strong spin-orbit interactions can also realize a 1D topological superconducting phase. In this setting, the subgap states propagating along the junction are tuned into a topological superconducting phase by controlling the phase bias across the junction and applying an in-plane magnetic field (Figs.\ 2D and 3D) \cite{Pientka2017,Hell2018}. The phase bias significantly enlarges the topological region of the phase diagram, where MZMs are confined at the ends of the junction. Crucially, this makes the topological phase quite robust to changes in geometry and chemical potential, although the supercurrent induced by the phase bias tends to diminish the gap.

Apart from these model platforms, numerous other ideas, based on a wide variety of materials and hybrid structures, have been proposed for creating topological superconductivity. Ideas for realizing topological qubits and Majorana-based quantum computations are most developed for the nanowire platform  (see Box 2). After briefly describing the current experimental status of detecting MZMs in the model platforms, we sketch the potential of other proposed approaches for advancing the field. Ultimately, besides creating topological qubits, the field aims at providing a platform for realizing and discovering new topological phases and phenomena. 
\medskip


\noindent\fbox{%
    \parbox{0.47\textwidth}{%
\begin{center}
\includegraphics[width=.7\linewidth]{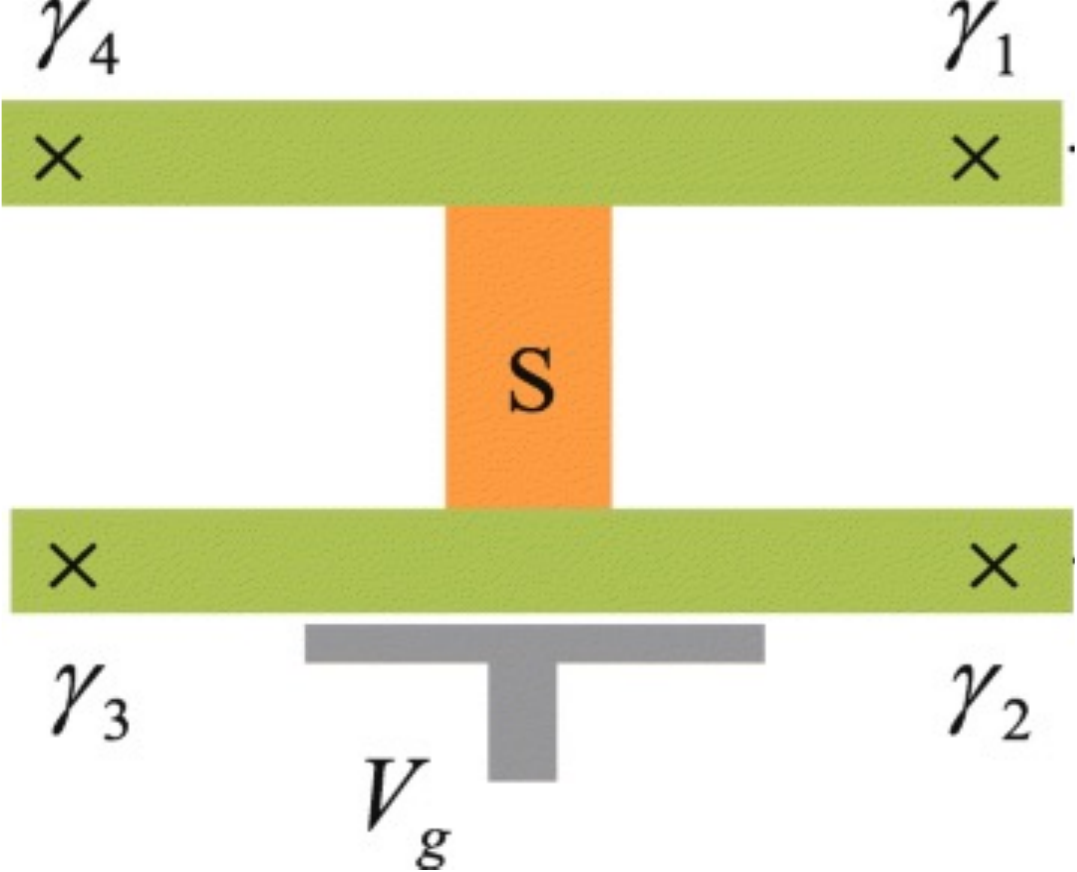}\end{center}

\textbf{Box 2: Majorana-based topological qubit} -- 
Ideas for realizing topological qubits are most developed for Majorana wires \cite{Karzig2017,Plugge2017}. The two degenerate ground states of a single Majorana wire have different fermion parities and cannot be brought into a coherent superposition, because of fermion-parity superselection. Coherent qubit dynamics requires at least two quantum wires (green) with four MZMs $\gamma_j$ ($j=1,\ldots,4$). Their ground-state manifold is spanned by four states, two with even and two with odd fermion parity. Both parity sectors can act as the two-level system of the topological qubit.
After connecting the two wires by a conventional superconductor (S, orange), the total electron number and hence the fermion parity can be controlled by a gate voltage $V_g$.  
Coulomb blockade can therefore help protect the qubit against leakage errors into the non-computational subspace. The Pauli operators of the qubit are bilinears in the Majoranas, e.g., $X=i\gamma_2\gamma_3$ and $Z = i\gamma_1\gamma_2$, which can be read out by appropriately coupling to the respective pair of MZMs \cite{Karzig2017,Plugge2017,Steiner2020}. This setting is known as a Majorana box qubit or tetron. A minimal test of a topological qubit would be the implementation of sequential Stern-Gerlach-type experiments.
\medskip

}}
\medskip

\begin{figure*}

\includegraphics[width=\linewidth]{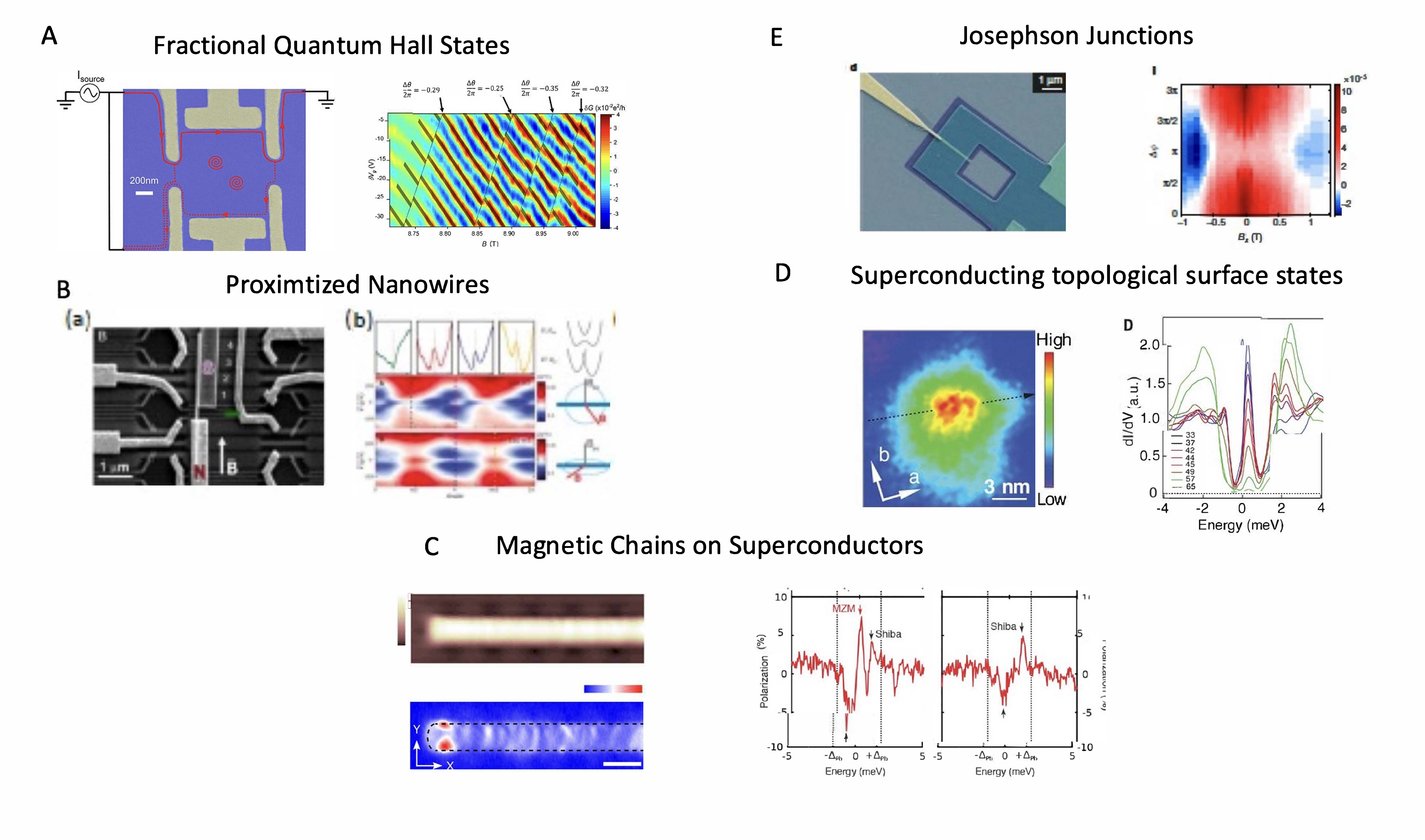}
\caption {\label{Fig3} Experimental platforms for MZM searches. A. Interferometry of FQH edge states can be used to probe the phase acquired in exchange processes of FQH quasi-particles. Such experiments have confirmed that the quasi-particles of the $\nu=1/3$ state are abelian anyons \cite{manfra20}. Similar experiments on candidate non-abelian states, such as the $\nu=5/2$ state, may provide evidence for Majoranas associated with non-abelian quasi-particles. B. Experiments on semiconductors with strong spin orbit coupling, proximitized by a superconductor have been at the focus of numerous works in search of 1D topological superconductivity and MZMs. The zero bias peak routinely observed in these experiments \cite{Mourik2012} appears for field orientations along the wire and has been interpreted as evidence for MZMs. However, other experiments show that such features may also originate from Andreev bound states. C. Chains of magnetic atoms on the surface of a superconductor provide another possible platform for realizing MZMs. Experiments on a ferromagnetic Fe chain on a Pb surface show spatially localized zero-energy states \cite{Feldman2017} with characteristic signatures of MZMs in spin-resolved measurements \cite{Jeon2017,He2014}. The large number of low energy in-gap states in such chains has made the identification of the expected topological gap difficult. D. Topological surface states in Fe-based superconductors have also been investigated as a possible platform for MZMs. Spectroscopy of  vortex cores on such surfaces reveal zero bias peaks with some reported behavior distinct from trivial vortex bound states \cite{Wang333}. The interplay between states on the surface and in the bulk of such vortices makes stablizing MZM in this setting sensitive to bulk properties.  E. Phase-biased Josephson junctions using strongly spin-orbit-coupled semiconductors in the presence of an in-plane magnetic field are predicted to exhibit a topological phase with MZMs. Experiments on such junctions with HgTe show zero bias peaks in tunneling spectroscopy \cite{Ren2019}. The difficulty in distinguishing zero bias peaks due to MZMs from other non-topological in-gap states across the various platforms calls for more sophisticated experiments.}
\end{figure*}

\subsection{Experimental Searches and Lessons Learned}

\textbf{Fractional quantum Hall states} 
The search for non-abelian quasi-particles in fractional quantum Hall systems has focused on detecting the quantum mechanical phase associated with quasi-particle exchanges in electronic interferometers \cite{an-inf1,H225}.
Recently, such experiments on a GaAs-based 2D electron gas measured the phase shift associated with the exchange of the abelian quasi-particles in the $\nu=1/3$ fractional quantum Hall state \cite{manfra20}. Similar experiments on the $\nu=5/2$ state of a GaAs device  found results  consistent with expectations for Ising anyons \cite{willett2019} (Fig.\ 3A) . However, so far the evidence remains statistical, and various questions remain about the microscopic geometry of the interference region. Several experiments confirmed that the quasi-particles have charge e/4 \cite{Bid2008,Venkatachalam2011}, but such charges could also occur, in principle, in a system with abelian statistics. 

As mentioned above,  three topologically distinct states, all containing Ising anyons, are competing candidates to explain the Hall plateau at $\nu=5/2$. Numerical studies of finite systems have strongly favored a state originally proposed by Moore and Read \cite{MooreR91}, known as the Pfaffian state, or perhaps preferably, its particle-hole (PH) conjugate, the anti-Pfaffian state.  However, measurements of the quantized thermal conductance of the edge modes at $\nu=5/2$  \cite{BanerjeeHUFOS18} have strongly favored a third state, known as the PH-Pfaffian, which has not appeared as a plausible  candidate in numerical studies. Further evidence for the PH-Pfaffian has come from measurements of thermal transport in geometries where the 5/2 state is bordered by states with $\nu=2$ and $\nu=3$ \cite{interface-5/2}. The resulting discrepancy between theory and experiment remains to be understood (cf., \cite{FeldmanHalperin21} and references therein). 

\textbf{One-dimensional systems}
In the search for experimental signatures of MZMs, much attention has focused on one-dimensional systems. The putative MZMs are commonly probed by tunneling from a normal-metal electrode \cite{Mourik2012}. At biases within the superconducting gap, tunneling is dominated by the hybridization of the electrode states with the MZM at zero energy, predicting a subgap tunneling resonance at zero voltage. Zero-bias peaks in the differential conductance that are consistent with this expectation are routinely observed in studies of proximity-coupled semiconductor nanowires with strong spin-orbit coupling and subject to a parallel field. (Fig.\ 3B) Numerous studies have examined whether this finding is consistent with a MZM interpretation \cite{LutchynReview}. Possible alternatives with similar phenomenology assume an Andreev bound state fine-tuned to zero energy by the suppression of pairing, the formation of multiple quasi-Majoranas, or the formation of a quantum dot near the end of the wire \cite{Kell2012,Prada2012,Lee2012,Lee2013,Liu2017,Vuik2019}. In a related experimental platform, the semiconductor nanowire is fully encapsulated by a superconducting shell (at the expense of a lack of gate control). In the presence of  magnetic flux threading the superconducting shell, zero bias peaks appear when  tunneling into the end of such a wire \cite{Vaitiekenas2020}; see also \cite{Thorp2021}. However, comparable phenomenology in other experiments on similar devices that explored a larger number of experimental configurations has been interpreted in terms of Andreev bound states. These can be tuned to zero energy by the magnetic flux \cite{Valentini2021}, providing an alternative explanation for the zero bias peaks.

Experiments on semiconductor nanowires in the Coulomb-blockade regime also provide an interesting probe of  MZMs \cite{Albrecht2016,Valentini2022}. Such experiments show a transition from $2e$ charging peaks at weak magnetic fields to $1e$ charging peaks at higher magnetic fields, consistent with the emergence of MZMs at the topological phase transition. In addition, the spacing of the Coulomb blockade peaks can be used to extract the residual splitting of the MZMs as a function of the length of the wire. An exponential length dependence of this splitting, reported in \cite{Albrecht2016}, was interpreted as consistent with the behavior of MZMs. A more recent study found a more complex picture \cite{Valentini2022}. For instance, the transition from $2e$ to $1e$-periodic Coulomb-blockade peaks was not accompanied by the emergence of zero bias peaks in tunneling into the ends of the nanowire. 

Under optimal conditions, a Majorana zero-bias peak is predicted to have a quantized height of $2e^2/h$ ($h$ is Planck's constant) \cite{Law2009,Flensberg2010,Wimmer2011,Pientka2012}. Experimental tests of the quantized conductance have been subject to considerable recent discussion. One report of its observation \cite{Zhang2018} was subsequently retracted \cite{Zhang2021,Frolov2021}, a second study exhibited scaling of the peak conductance that was interpreted as consistent with quantization  \cite{Nichele2017}. A pertinent complication is that experiments consistently observe a substantial softening of the superconducting gap, once the magnetic field is sufficiently strong to induce the putative topological superconducting phase. Furthermore, theorectial work suggests that peak quantization is not a definite indicator of Majorana end states, but can also occur outside of the topological superconducting phase \cite{Vuik2019, Pan2020}

Current efforts on the nanowire platform focus on cross-correlating the existence of zero-bias peaks with other signatures such as the gap closing and reopening at the topological phase transition \cite{Pikulin2021,Banerjee2022}. 
A recent experiment has implemented a previously defined gap protocol \cite{Pikulin2021}, designed to identify parameter regions, in which the nanowires  exhibit a topological superconducting gap with high likelihood. The experiment on
gate-defined 1D channels in hybrid InAs-Al devices
provided evidence of the expected cross correlations for a few devices and narrow ranges of chemical potential
\cite{Microsoft2022}.  

An important issue for the nanowire platform, but also more generally, is the role of disorder.
Unlike the case of $s$-wave superconductivity, which is largely immune to non-magnetic disorder, in odd-parity superconductors even weak disorder can have detrimental effects. In semiconducting wires proximitized by an $s$-wave superconductor, potential disorder introduces subgap states that may considerably soften the gap, lead to partitioning of the wire into segments each having localized Majorana zero modes, and drive the system out of the topological superconducting phase beyond a critical strength of disorder \cite{Motrunich2001,Brouwer2011}. 
Disorder also induces large fluctuations of the residual Majorana splitting \cite{Brouwer2011b} and can result in experimental signatures such as zero bias peaks in conductance that are not indicative of Majorana zero modes \cite{Bagrets2012}. Considerable effort has gone over the years into simulating the effects of disorder in more realistic models of the nanowire platform, see, e.g., \cite{Sau2012,Ahn2021, Pikulin2021,Microsoft2022}. It is also interesting to comment that although disorder in the nanowire is potentially detrimental, disorder in the superconducting shell can be crucial to establish effective proximity coupling \cite{Reeg2018,Kiendl2019}.  

In chains of magnetic atoms, spectroscopy with a STM has detected signatures of highly localized zero-energy modes at the ends of ferromagnetic chains (Fe) assembled on the surface of a strongly spin-orbit-coupled superconductor (Pb) \cite{NadjPerge2014,Ruby2015,Pawlak2016}. Just as for the nanowire platform, zero-bias peaks by themselves do not distinguish between MZMs and conventional Andreev bound states or quasi-Majoranas. However, unlike conductance measurements on device-like structures, STM experiments can spatially resolve MZMs and other low-energy states, allowing one to distinguish between end and bulk states. The zero-bias peaks detected in these experiments at the ends of the chain are robust to being buried with an additional superconducting layer \cite{Feldman2017} and display spin-polarization signatures that are consistent with a MZM (Fig.\ 3C)  \cite{Jeon2017,He2014}.  The spin-polarized measurements on the Fe chains also provide evidence against impurity-induced end  states. However, they cannot eliminate the possibility of multiple MZMs originating from different channels, if their splitting is below the experimental resolution (currently at 80 $\mu$eV). STM spectroscopy of chains that exhibit signatures of zero-energy end states also show residual in-gap states along the chain, obstructing the identification of a topological bulk gap \cite{Feldman2017}. Theoretical studies have addressed  mechanisms for the appearance of trivial zero-energy end states \cite{Sau2015,Hess2022}.

Several experiments have explored possible signatures of MZMs in magnetic atom chains with various combinations of magnetic adatoms and substrates, including chains assembled atom by atom \cite{Ruby2017,Kim2018,Kamlapure2018,Schneider2021,Liebhaber2022,Schneider2022}. These experiments highlight that the emergence of zero-energy modes localized near the chain ends is not a generic feature of atomic chains. Some experiments do not observe zero-energy features at all, whereas they are not localized at the ends of the chains in others. This is consistent with the theoretical expectation that the emergence of topological superconductivity is contingent on conditions such as the number of 1D bands (related to the occupation of the adatom $d$-shell), the strength of spin orbit coupling, or the magnetic ordering. In dilute chains, topological superconductivity may also be suppressed by the quantum nature of the adatom spins \cite{Steiner2022,Liebhaber2022}.

\textbf{Topological boundary modes and vortex cores} Efforts to realize MZMs in topological boundary modes initially focused on heterostructures of topological insulators and conventional superconductors. STM spectroscopy was used to probe properties of vortex cores in the resulting superconducting 2D surface states.  We note that in these settings, the presence or absence of MZMs depends sensitively on doping as the vortex core must realize a one-dimensional topological superconducting state to support MZMs at its ends \cite{Hosur2011,Chiu2011}.

More recent efforts investigate Fe-based superconductors, which combine a 2D topological surface state with intrinsic bulk superconductivity. STM experiments on the surface of FeTe$_{0.55}$Se$_{0.45}$ reported the observation of a sharp zero-bias peak inside the vortex core \cite{Wang333,Kong2019}. Although some of the reported properties in these measurements are consistent with MZMs, distinguishing them from trivial Caroli–de Gennes–Matricon  states remains challenging \cite{Jack2021}. For example, only a fraction of vortices were actually found to host a zero-energy state, with others hosting Caroli–de Gennes–Matricon states at finite energy. The role of surface disorder in these experiments has also added challenges. Experiments have reported that disorder can induce trivial surface states in FeTe$_{0.55}$Se$_{0.45}$ and that vortex-core zero-energy states are even entirely absent \cite{Kong2019,Chen2019b}. Reports of plateau-like behavior in STS measurements of zero-energy vortex-core states as a  function of tip-sample distance have been been interpreted as a possible observation of the quantized conductance for tunneling into a MZM \cite{Zhu2020}. However, similar experiments show larger conductance than the predicted quantized value\cite{Chen2019q}, and the plateau-like behavior may be caused by suppressed quasiparticle relaxation at higher tunneling current into a trivial vortex core state, rather than by quantization \cite{Jack2021}.

Interesting high-resolution STM experiments of vortex cores in FeSe$_{0.5}$Te$_{0.5}$ show a zero-bias resonance that is separated from finite-energy resonances and occurs only at low fields, presumably owing to core-core overlap of the zero modes at higher magnetic fields (Fig.\ 3D) \cite{Machida2019}. These most recent experiments highlight the key advantage of Fe-based superconductors, in which the short coherence length can potentially allow for distinguishing Caroli–de Gennes–Matricon states from MZM.  Theoretical work has discussed the possible emergence of trivial zero-energy states due to impurities \cite{Mendonca22}.

Experiments have also explored inducing superconductivity and magnetism in the topological edge modes of Bi bilayers \cite{Jack2019}. The Bi bilayers are grown on a superconductor and their edges are covered by magnetic clusters. Probing the edge mode by local STM spectroscopy, experiments detect zero-bias peaks between regions influenced by superconductivity and magnetism, respectively. Similar to the experiments on atomic chains \cite{Jeon2017}, spin-polarized STM was used to show that the zero-energy states have spin properties distinct from those induced by magnetic clusters and consistent with those expected for a MZM. 

\textbf{Josephson junctions} Signatures of topological superconductivity have also been observed in Josephson junctions with the strongly spin-orbit-coupled semiconductors HgTe and InAs (Fig.\ 3E) \cite{Fornieri2019,Ren2019}.
The application of a phase bias across the junction enlarges the topological region. At the same time, the gap is reduced by the in-plane magnetic field $\sim $ 1 T (weakening superconductivity in the electrodes) and by the supercurrent associated by the phase bias. Moreover, the gap can become soft due to electron and hole trajectories which propagate nearly in parallel to the superconducting electrodes and are thus weakly affected by the superconductor. In experiment, this reduction results in a large localization length ($\sim \mu$m) of the MZMs, obscuring their experimental signatures and complicating implementations of braiding. 
\textbf{Other efforts} Metallic surface states can have particularly strong Rashba spin-orbit coupling and thus provide a possible alternative to spin-orbit-coupled semiconductors for realizing MZMs \cite{Potter2012}. This has stimulated the proposal to engineer a (quasi-)1D topological superconductor using Au nanowires on a superconducting substrate, despite their much smaller Fermi wavelength. Recent STM experiments on Au wires proximity coupled to the ferromagnetic insulator EuS show preliminary signatures consistent with MZMs \cite{Manna2019}. However, structural variations of the samples have thus far hampered consistent observations of MZM signatures, and high-resolution experiments would be desirable. 

\begin{figure*}
\includegraphics[width=\linewidth]{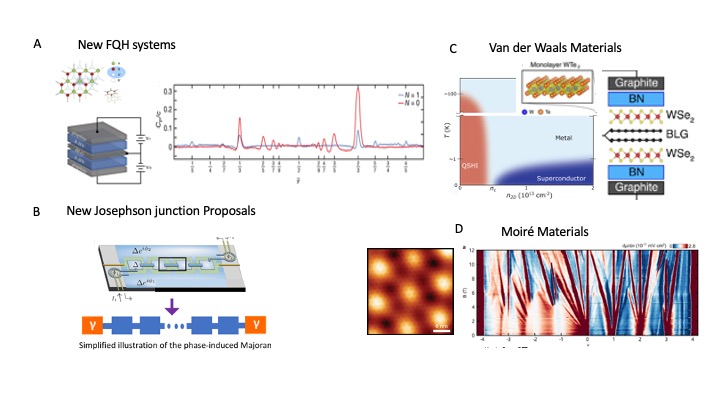}
\caption {\label{Fig5} Future directions. A number of recent developments are expanding the material platforms and device concepts for realizing  MZMs and other non-abelian quasi-particles. A. A graphene bilayer encapsulated by boron nitride offers a  promising platform for pursuing experiments on even denominator FQH states, detected here by  capacitance measurements
as a function of electron density and perpendicular displacement field \cite{gr4,gr5}. B. Chain of Josephson junctions with multiple superconducting islands. Controlling their superconducting phases is predicted to allow for realizing more localized MZMs \cite{Lesser2021b}. C. The rapidly expanding field of 2D van der Waals materials is creating  opportunities for future research on MZMs. The discovery of both, a topological insulating phase and superconductivity in gated  WTe$_2$ devices has opened up the possibility of laterally combining these phases within the same material \cite{Fatemi2018} and thereby   creating the conditions for MZMs. Similarly, inducing spin-orbit coupling into graphene-based superconducting systems by proximity \cite{Island2019} can be used to create new platforms for MZMs. D. Moire materials have dramatically expanded discovery of superconducting and topological phases, including interaction-driven Chern and fractional Chern insulating phases \cite{Nuckolls021,Xie2021}. Compressibility measurements on magic-angle graphene 
as a function of carrier density and magnetic field provide a highly sensitive probe of such phases \cite{Xie2021}.}
\end{figure*}

\subsection{Future directions}

The search for non-abelian anyons continues to provide ample opportunities for experiments across a wide range of material platforms and for exploration of a broad spectrum of quantum phenomena. The immediate challenge in the field remains to firmly establish the experimental existence of MZMs and other non-abelian anyons. It is now clear that by themselves, zero-bias peaks do not reliably indicate MZMs. Even the recent minimal approach of cross-correlating several features of the topological phases \cite{Pikulin2021,Banerjee2022,Microsoft2022} remains largely insensitive to quasi-particle poisoning and therefore oblivious to the topological ground-state degeneracy associated with the presence of MZMs. Both in superconducting platforms and in FQH systems, experimental tests of the novel fusion properties would be highly desirable to establish the non-abelian behavior of these quasi-particles. More stringent, next-generation tests of MZMs would also probe the coherent dynamics within their degenerate subspace (see Fig.4). This constitutes an essential precursor to implementing topological qubits and to full-scale tests of non-abelian statistics.

\textbf{Quantum Hall systems} A major goal for studies of fractional quantum Hall systems is to find more convincing evidence for non-abelian quasi-particles. It would be desirable to improve measuring techniques for interference experiments on quasi-particles traveling along edge states. It would also be helpful to obtain better control of the electron gas in GaAs devices, for instance to control more precisely when a localized quasi-particle enters or leaves the area enclosed by the interferometer. It is also important to pin down the actual nature of the states at filling factor $\nu=5/2$ and $7/2$, and to resolve the current disagreements between theory and experiment.

An exciting new direction for exploring non-abelian anyons is to examine the even-denominator FQH states observed in systems other than GaAs \cite{ZnO2,gr1,gr2,gr3,gr4,gr5}. Bilayer graphene constitutes a particularly promising system \cite{gr4,gr5}. When encapsulated by hexagonal boron nitride, it has a smaller dielectric constant ($\sim 4.4$) than GaAs systems ($\sim 12.8$), implying stronger Coulomb interactions. Consequently, experiments on bilayer graphene detected even-denominator states with gaps that are an order of magnitude larger (5K at 14T) than in GaAs. Recent experiments indicate that these even-denominator states are dominantly spin and valley polarized, ruling out competing abelian ground states \cite{gr5} (Fig.\ 5A). Combining 2D layered materials and their stacks with new nanofabrication techniques, interference as well as fusion experiments with trapped non-abelian anyons may become possible. Recent advances in STM spectroscopy of FQH states \cite{XLi2021} may also be used to detect non-abelian anyons trapped near defects \cite{stm}. 

Quasi-particles with richer non-abelian statistics than that of Ising anyons would ultimately be needed for universal quantum computation with full topological protection. Suitable excitations 
have been predicted for certain quantized Hall states, such as the $\nu=12/5$ state in GaAs structures \cite{MZ-12/5,non-AB-12/5,non-AB-12/5-RR,BS-12/5,non-AB-12/5-LLM,non-AB-12/5-LLM-2,non-AB-12/5-ZGSH}. However, corresponding experimental evidence is not strong (ref?). Theoretical work has also shown that states supporting various types of non-abelian quasi-particles can be produced in hybrid structures of abelian quantum Hall systems and superconductors \cite{Lindner2012,Clarke2013,Mong2014}. These proposals have so far not been realized in experiment as superconductivity and quantum Hall systems have conflicting magnetic-field requirements. Moreover, the anyons produced in the structures proposed in \cite{Lindner2012} and \cite{Clarke2013} would still fall short of what is needed for universal topologically protected quantum computing. The proposal of \cite {Mong2014} realizes anyons which allow universal topological quantum computation, but is rather challenging to implement experimentally.

Nevertheless, preliminary experiments, in which a narrow superconducting finger penetrates the edge of a Hall bar, have been successfully realized and crossed Andreev processes have been detected \cite{GHLee2017}. In this geometry, the finger can be thought of as bending the chiral edge modes along its boundary, thereby forming two counter-propagating modes proximity-coupled to a superconductor. If the edge channels are spin polarized and the superconductor is intrinsically spin-orbit coupled to facilitate pairing, this setup might support topological superconductivity. The problem of large magnetic fields can also be circumvented by using thin films of magnetically doped topological insulators (TI), which exhibit the quantum anomalous Hall (QAH) effect even in the absence of a magnetic field.

Perhaps even more intriguing is the possibility of proximity-coupling fractional Chern insulating states and superconductors \cite{Gul2020}. Theory predicts that such hybrid systems can also host more complex excitations, known as parafermions, with  richer braiding statistics than Ising anyons \cite{Alicea2016}. Realizing fractional Chern insulators in conditions compatible with superconductivity is most likely in new materials systems, such as the moir\'e materials considered below.

\textbf{Josephson junctions} Several  approaches have been proposed to improve the existing experiments searching for topological superconductivity in Josephson junctions. This platform exploits the fact that applied supercurrents reduce the threshold magnetic field required to enter the topological phase \cite{Romito2012}. As shown recently, supercurrents can obviate the need for a magnetic field entirely, with the applied phase difference across the junction providing the required breaking of time reversal symmetry \cite{Melo2019,Lesser2021,Lesser2021b,Lesser2022b}. One approach inserts a chain of additional superconducting islands into the original Josephson junction geometry. For appropriate phase biasing of the islands relative to the superconducting banks of the Josephson junction, each unit cell containing a single island harbors two MZMs (Fig.\ 5B). Neighboring MZMs belonging to  different unit cells gap out, leaving only the isolated MZMs at the ends of the junction intact. In this setting, the additional phase knob introduced by the intermediate islands enables the realization of a topological superconducting phase in the absence of a magnetic field. The localization length of the MZMs is set by the size of the unit cell, so that the MZMs can be effectively separated further by adding more unit cells along the wire. Counter-intuitively, it has been pointed out that disorder can stabilize the topological phase in this platform
\cite{Haim2019b,Laeven2020}.

\textbf{1D topological superconductors}
Work to date has focused on MZMs in systems with broken time reversal symmetry. Time reversal symmetric topological superconductors support stable Kramers pairs of MZMs with fractional boundary spin \cite{Haim2019}. Advances in creating more robust 1D topological superconductors in hybrid structures would open exciting possibilities for engineering more complex phases with novel topological properties and for exploring exciting quantum phenomena. Josephson junctions of 1D topological superconductors should exhibit the celebrated $4\pi$-periodic Josephson effect. Coupling several 1D topological superconductors to a disordered quantum dot has been predicted to provide a realization of the Sachdev-Ye-Kitaev (SYK) model \cite{Pikulin2017,SerosAlicea2017}. Signatures of this phase can be observed by tunneling into the quantum dot. Arrays of 1D systems with controllable coupling between them provide a pathway to realizing a 2D topological superconducting phase with multiple chiral and non-chiral modes at its boundary \cite{Seroussi2014}.  

Coulomb blockaded islands with multiple MZMs are particularly powerful building blocks in the context of topological quantum computing \cite{OregOppenAnnualReview}. Islands with four MZMs, known as tetrons \cite{Karzig2017} or Majorana box qubits \cite{Plugge2017}, effectively implement a topological qubit or, equivalently, local spin degrees of freedom (see Box 2). Hybridization of MZMs between islands realizes anisotropic spin-spin interactions or even higher spin-cluster interactions. This can be exploited to faithfully realize the Hamiltonians of topological error correcting codes \cite{Vijay2015,Landau2016}. While such networks of coupled islands of topological superconductors realize 2D topological superconductors in the absence of charging effects, charging can effectively drive them into topologically ordered insulating states. 

\subsection{New Material Platforms} 
Beyond improvements to existing material platforms, the discovery of novel materials promises to play a central role in advancing the field of topological quantum phenomena and non-abelian anyons. The potential of new materials is exemplified by the discovery of tunable superconductivity and topological phases in monolayers of WTe$_2$ (Fig.\ 5C) \cite{Fatemi2018}. WTe$_2$ combines superconductivity with strong intrinsic spin-orbit coupling as well as gate tunability into a 2D topological-insulator phase. This material is thus a promising platform for realizing phase controlled topological superconductivity without the need to induce superconductivity externally.

Robust and tunable superconductivity has also been observed in twisted matter such as magic angle twisted bilayer, trilayer, and multilayer graphene \cite{Cao2018,Hao2021,Zhang2022,Park2022}. Although graphene has very weak intrinsic spin-orbit coupling, spin-orbit coupling can be induced by including, e.g., additional WSe$_2$ layers into the device \cite{Island2019} (Fig.\ 5C). These systems have been shown to support Chern insulator (FCI) states down to low magnetic fields \cite{Nuckolls021}. Most recently, FCIs have been seen at magnetic fields as low as 5T \cite{Xie2021} (Fig.\ 5D). In present samples, other phases are stabilized at lower magnetic fields. Understanding the competition between these phases and what physical parameters favor the FCI state will help reduce the magnetic field in which FCIs can form such that both superconductivity and FCI phases can coexist in the same sample. 

These systems could thus provide an interesting platform to search for parafermions, combining low magnetic fields with intrinsic superconductivity. Proximity coupling various van der Waals materials with magnetic and superconducting ground states may also be used to create hybrid structures for topological superconductivity. It would be particularly exciting if non-abelian fractional quantum Hall states could be realized in ferromagnetic structures without an applied magnetic field. So far, there is evidence for zero-field integer anomalous quantum Hall states in several materials \cite{Chang15,Bestwick15}, but fractional states have not yet been observed.

There have also been efforts to combine various magnetic materials, including doped topological insulators, with superconductors to realize chiral Majorana edge modes. These works are motivated by  proposals to employ chiral edge modes as a platform for topological quantum computing \cite{Lian2018}. However, this area is currently far less advanced experimentally and initial results \cite{He2017} have been retracted (ref?) and shown to lack reproducibility \cite{Kayyalha2020}. Chiral Majorana edge modes also appear in the exactly solvable Kitaev honeycomb model, a spin model with anisotropic exchange couplings \cite{KITAEV20062}. Experimental realizations have been proposed based on judiciously engineering the hybridization of Coulomb-blockaded islands hosting multiple MZMs \cite{Thomson2018,Sagi2019,OregOppenAnnualReview}. Realizations in Kitaev materials \cite{Jackeli2009,Trebst2017}, possibly evidenced by thermal transport experiments \cite{Kasahara2018,Vinkler2018,Ye2018}, have also been discussed in the context of Majorana-based quantum computation \cite{Aasen2020}.

Exploiting the steady advances of quantum simulations with existing quantum computing platforms, there have also been interesting efforts \cite{Mi2022} to simulate Floquet incarnations of spin models such as the quantum Ising model, which is closely related to the Kitaev chain. These models also support zero (as well as $\pi$) modes and provide a setting for studying their stability to perturbations in chains of superconducting qubits. This arena may help explore the utility of topological concepts such as Majorana zero modes in advancing noisy intermediate quantum technologies beyond their current coherence limitations.

In conclusion, experimentally establishing the existence of non-abelian anyons such as Majorana zero modes constitutes an outstandingly worthwhile goal, first from the point of view of fundamental physics, but also due to their potential applications. Triggered by the theoretical understanding of non-abelian fractional quantum Hall states and superconductor-topological insulator hybrids, there have been numerous attempts to observe non-abelian anyons in the laboratory. Many claims are based on rather circumstantial evidence and, apart from a few extensively studied platforms, were not subjected to intense scrutiny or in-depth analysis of alternative interpretations. Future progress will be possible when claims of Majorana discoveries are based on experimental tests that go significantly beyond indicators such as zero-bias peaks, which at best suggest consistency with a Majorana interpretation. It will be equally essential that they build on an excellent  understanding of the underlying materials system. It seems likely that further material improvements of existing platforms and the exploration of new material platforms will both be important avenues to make progress towards 
solid evidence for Majoranas. Then we can hope to explore -- and harness -- the fascinating physics of the topologically protected ground state manifold and nonabelian statistics.

\subsection{Acknowledgment}
We are grateful to all our collaborators on this subject. A.\ Yazdani acknowledges funding from Gordon and Betty Moore Foundation’s EPiQS initiative grant GBMF9469, ONR grant N00014-21-1-2592, NSF-MRSEC through the Princeton Center for Complex Materials grant NSF-DMR-2011750, the U.S.\ Army Research Office (ARO) MURI project under grant number W911NF-21-2-0147, and NSF grant DMR-1904442.  BIH acknowledges funding from the Microsoft Corporation and NSF grant DMR-1231319. FvO is grateful for funding by the Deutsche Forschungsgemeinschaft (CRC 183 and SFB 910) as well as the Einstein Research Unit on Quantum Devices.
A.\ Yacoby is supported by the Quantum Science Center (QSC), a National Quantum Information Science Research Center of the U.S. Department of Energy (DOE). A.\ Yacoby is also partly supported by the Gordon and Betty Moore Foundation through Grant GBMF 9468, by the U.S.\ Army Research Office (ARO) MURI project under grant number W911NF-21-2-0147, and by the STC Center for Integrated Quantum Materials NSF Grant No.\ DMR-1231319.

\bibliographystyle{apsrev4-1}
\bibliography{NatPerspBib}

\end{document}